\documentclass[twocolumn,twoside]{revtex4}
\usepackage{graphicx}
\usepackage{fancyhdr}
\usepackage{slashed}
\newcommand{\mathsym}[1]{{}}

\newcommand{\baz}{\begin{array}{cc}}
\newcommand{\bad}{\begin{array}{ccc}}
\newcommand{\ba}{\begin{array}{c}}
\newcommand{\ea}{\end{array}}
\newcommand{\be}{\begin{equation}}
\newcommand{\ee}{\end{equation}}
\newcommand{\bea}{\begin{eqnarray}}
\newcommand{\eea}{\end{eqnarray}}

\newcommand{\bi}{\begin{itemize}}
\newcommand{\ei}{\end{itemize}}
\newcommand{\bmt}{\begin{pmatrix}}
\newcommand{\emt}{\end{pmatrix}}
\newcommand{\bt}{\begin{tabular}}
\newcommand{\et}{\end{tabular}}

\newcommand{\benu}{\begin{enumerate}}
\newcommand{\eenu}{\end{enumerate}}

\newcommand{\mwl}{m_{W_L}}

\newcommand{\muegam}{\mu\to e\gamma}

\newcommand{\bav}{\begin{array}{cccc}}

\begin{document}

\title{Determining Neutrino mass hierarchy in an extended Left-Right Model}

\author{Prativa Pritimita$^*$}
\affiliation{Indian Institute of Technology Bombay, Mumbai, India}
\author{Nitali Dash}
\affiliation{School of Physics, Regional Institute of Education (NCERT), Bhubaneswar, India}
\author{Sudhanwa Patra}
\affiliation{Indian Institute of Technology Bhilai, Chhatisgarh, India}
\author{Urjit A. Yajnik}
\affiliation{Indian Institute of Technology Bombay, Mumbai, India}

\begin{abstract}
We derive the lower bound on absolute scale of lightest neutrino mass for normal 
hierarchy and inverted hierarchy pattern of light neutrinos by studying the new 
physics contributions to charged lepton flavour violating decays and neutrinoless double beta decay in the framework of 
a TeV scale left-right symmetric model. Neutrino mass is generated in the model via type-II seesaw dominance with the help of a heavy sterile neutrino. This scenario allows large light-heavy neutrino mixing and the mixing facilitates new channels for neutrinoless double beta decay and lepton flavour violating decays. We express all the model parameters in terms of oscillation parameters in order to constrain light neutrino mass scale and extract information on neutrino mass hierarchy.
\end{abstract}

\maketitle

\thispagestyle{fancy}

\section{Introduction}
Left-Right Symmetric Model (LRSM) ~\cite{Mohapatra:1974gc,Pati:1974yy,Senjanovic:1975rk,Senjanovic:1978ev} was first proposed to explain parity restoration at high scale. The theory can also explain the origin of neutrino mass through seesaw mechanism since it has a right-handed neutrino. Many forms of LRSM are possible. In generic LRSM, the symmetry breakings happen with the help of scalar triplets and bidoublet which leads to the generation of neutrino mass through type-I plus type-II seesaw mechanism. 
\begin{equation}
M_\nu \ = \ -M_DM_R^{-1}M_D^T + M_L \ \equiv \ M_\nu^{\rm I}+M_\nu^{\rm II}  \nonumber
\label{eq:1}
\end{equation}
Here $M_D$ is the Dirac neutrino mass induced by the bidoublet vacuum expectation value (VEV), while $M_R$ and $M_L$ are the Majorana masses of the right and left-handed neutrinos respectively induced by the triplet VEVs.
For phenomenological purposes, usually one of the contributions is assumed to be dominant with observable ramifications for different experiments. For type-I seesaw dominance, it is assumed $M_L\to 0$ and the light neutrino mass depends on the Dirac mass matrix $M_D$. However, in this case all the light-heavy neutrino mixing effects are suppressed for TeV-scale parity restoration. Type-II seesaw dominance can be realized either by assuming $M_D\to 0$ or with very high scale of parity restoration. Here the light and heavy neutrino mass matrices are directly proportional to each other and lead to a more predictive scenario. But in this case also, the light-heavy neutrino mixing effects on lepton number or lepton flavor violating observables are lost. 

We have extended the generic LRSM only in the fermion sector by adding a sterile neutrino $S_L$ per generation which helps in exactly cancelling the type-I seesaw term and achieving natural type-II seesaw dominance \cite{Pritimita:2016fgr,Dash:2021pbx}. This scenario also allows large light-heavy neutrino mixing which has important and non-trivial phenomenological consequences. This mixing in the neutrino sector facilitates new channels for lepton number and lepton flavour violating decays. We have expressed these new physics contributions to cLFV and $0\nu\beta\beta$ decay in terms of the observed light neutrino oscillation parameters and lightest neutrino mass. As a result of this, we can derive constraints on the lightest neutrino mass from the non-observation of these rare processes. Even though many new channels are possible for these decays in the model, in order to highlight the contributions of right-handed heavy neutrino and sterile neutrino to LFV decays and $0\nu\beta\beta$ decay, we have focussed only on the diagrams mediated by them and ignored other possible channels.

\section{The Model Framework}

The left-right symmetric theory is based on the gauge group $\mathcal{G}_{LR} \equiv SU(3)_c\times SU(2)_L \times SU(2)_R \times U(1)_{B-L}$. The particle content of our model can be found in Table.\ref{tab:1}. The fermion sector contains all the usual quarks and leptons plus one extra sterile neutrino $S_L$. The scalar sector contains the doublets $H_L$, $H_R$, triplets $\Delta_L, \Delta_R$ and bidoublet $\Phi$. In the model, both doublets and triplets are used for the left-right symmetry breaking.

\begin{table}
	\begin{center}
	\caption{Particle content of the model}
	\vspace{5pt}
	\label{tab:1}
		\begin{tabular} {|c|c|c|c|c|c|}\hline
			& Fields & $SU(3)_c$ & $SU(2)_L$ & $SU(2)_R$ & $B-L$ \\ \hline
	& $q_L$ & {\bf 3} & {\bf 2} & {\bf 1} & 1/3\\
			& $q_R$ & {\bf 3} & {\bf 1} & {\bf 2} & 1/3\\
			& $\ell_L$ & {\bf 1} & {\bf 2} & {\bf 1} & $-1$ \\
			& $\ell_R$ & {\bf 1} & {\bf 1} & {\bf 2} & $-1$ \\
			& $S$ & {\bf 1} & {\bf 1} & {\bf 1} & 0\\
			\hline
	& $\Phi$ & {\bf 1} & {\bf 2} & {\bf 2} & 0 \\
			& $H_L$ & {\bf 1} & {\bf 2} & {\bf 1} & $-1$ \\
			& $H_R$ & {\bf 1} & {\bf 1} & {\bf 2} &  $-1$ \\
			& $\Delta_L$ & {\bf 1} & {\bf 3} & {\bf 1} & 2 \\
			& $\Delta_R$ & {\bf 1} & {\bf 1} & {\bf 3} &  2 \\
			\hline
		\end{tabular}
	\end{center}
	\end{table}
With these fermions and scalars, the interaction lagrangian for leptons can be written as,
\begin{eqnarray}
-\mathcal{L}_{Yuk} = \overline{\ell_{L}} \left[Y_3 \Phi + Y_4 \widetilde{\Phi} \right] \ell_R
+ f \left[\overline{(\ell_{L})^c} \ell_{L} \Delta_L+\overline{(\ell_{R})^c}\ell_{R}\Delta_R\right] \nonumber \\
+F \overline{(\ell_{R})} H_R S^c_L + F^\prime \overline{(\ell_{L})} H_L S_L + \mu_S \overline{S^c_L} S_L + \mbox{h.c.}
\label{lepton-interaction}
\end{eqnarray}
After spontaneous symmetry breaking, the neutral lepton mass matrix in the basis $\left(\nu_L, N^c_R, S_L\right)$ can be written as,
\begin{equation}
\mathcal{M}_\nu= \left( \begin{array}{ccc}
M_L                   & M_D   & 0 \\
M^T_D                 & M_R   & M^T \\
0          & M     & 0
\end{array} \right) \, ,
\label{eqn:numatrix}       
\end{equation}
where $M_D$ is the Dirac neutrino mass matrix connecting $\nu_L$ and $N_R$, $M$ is the mixing matrix in the $N_R-S_L$ sector, 
$M_L$ $(M_R)$ is the Majorana mass matrix for left-handed (right-handed) active neutrinos generated dynamically by non-zero 
VEV of scalar triplet $\Delta_L$ $(\Delta_R)$. Assuming the mass hierarchy, $M_R \gg M > M_D \gg M_L$, the light and heavy neutrino mass formulae can be written as, 
\begin{eqnarray}
 m_\nu = M_L \\
 m_N = M_R \\
 m_{S} \simeq M M^{-1}_R M^T
\end{eqnarray}

For detail derivation of neutrino masses and mixing in the model, please refer \cite{Dash:2021pbx}.

\section{LFV in LRSM}
In our model, LFV decays can be mediated by heavy right-handed neutrinos $N_R$, sterile neutrinos 
$S_L$, charged scalar triplets $\Delta^{\pm \pm}_{L,R}, \Delta^{\pm}_{L,R}$ due to purely left-handed currents involving $W_L$ and due to purely right-handed currents involving $W_R$.
 However we focus only on those contributions which involve large active-sterile neutrino mixing, i.e. due to the neutrinos $N_R$ and $S_L$ in order to constrain light neutrino masses from LFV decays. We ignore other contributions by imposing the limiting conditions on masses of $W_R$ boson and scalar trilpets $\Delta_{L,R}$ as, $M_{W_R} \gg M_{W_L}$ and $M_{\Delta_{L,R}}, M_{H_1} \gg M_{N,S}$.
 
\begin{figure}[h!]
	\centering
	\includegraphics[width=0.45\textwidth]{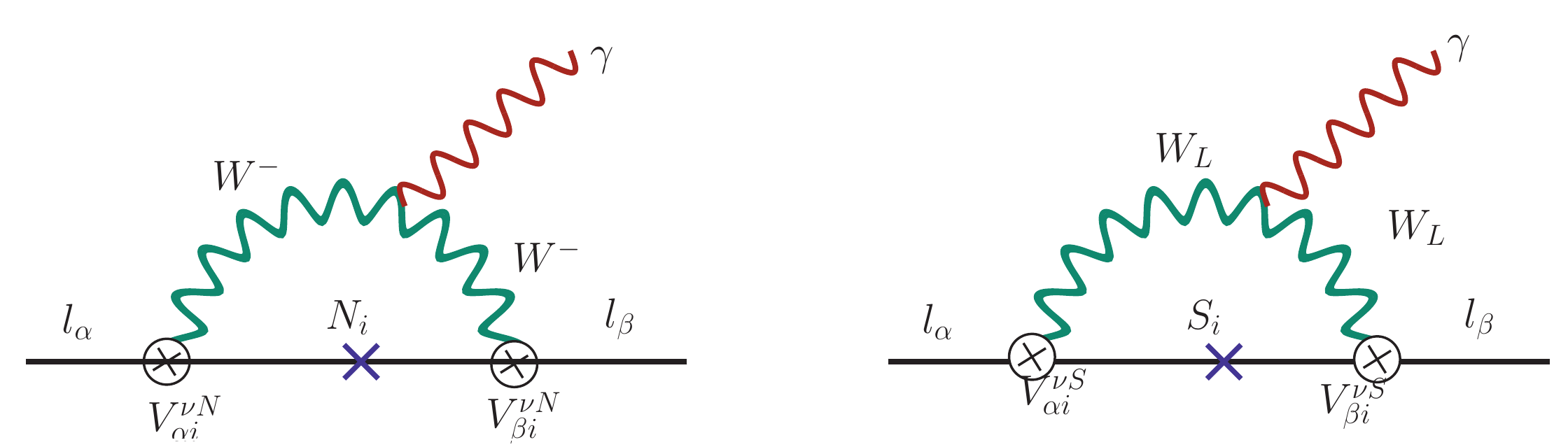}
	\caption{Feynman diagram for lepton flavor violating process $\muegam$ due to the exchange of mass eigenstates of heavy neutrinos $N_i$ and $S_i$.}
	\label{fig:lfv-NS}
\end{figure}
The effective Lagrangian relevant for the lepton flavor violating process $\mu \to e \gamma$ in our present work can be expressed as,
\begin{eqnarray}
\mathcal{L}_{\mu\rightarrow e} = -
\frac{e g^2}{4(4\pi)^2\mwl^2}m_\mu\overline{e}\sigma_{\mu\nu}(\mathcal{G}^\gamma_LP_L+\mathcal{G}^\gamma_RP_R)\mu F^{\mu\nu} \nonumber \\
-\frac{\alpha_W^2}{2\mwl^2}\sum_q \left\{\overline{e}\gamma_\mu\left[\mathcal{W}^q_LP_L
+\mathcal{W}^q_RP_R\right]\mu\; \overline{q}\gamma^\mu q\right\} + {\rm h.c.} 
\end{eqnarray}
where $\sigma_{\mu\nu} \equiv \frac{i}{2}[\gamma_\mu,\gamma_\nu]$ and the form factors 
are $\mathcal{G}^\gamma_{L,R}$ and $\mathcal{W}^{u,d}_{L,R}$. The relevant contributions 
to $\muegam$ is given by
\begin{eqnarray}
i\mathcal{M}(\mu \rightarrow e\gamma) = \frac{e\alpha_W}{8\pi\mwl^2}\epsilon_\gamma^\mu \overline{e} \mu
\left[\left(q^2\gamma_\mu - q_\mu\slashed{q}\right) \right. \nonumber \\ 
 \left.
\left(\mathcal{F}^\gamma_L P_L+\mathcal{F}^\gamma_R P_R\right) -im_\mu\sigma_{\mu \nu}q^\nu\left(\mathcal{G}^\gamma_L P_L+\mathcal{G}^\gamma_R P_R\right)\right]
\end{eqnarray}
where $\mathcal{F}^{\gamma}_{L,R}$ and 
$\mathcal{G}^\gamma_{L,R}$ are the anapole and dipole form factors. 
\begin{figure}[h!]
	\centering
	\includegraphics[width=0.38\textwidth]{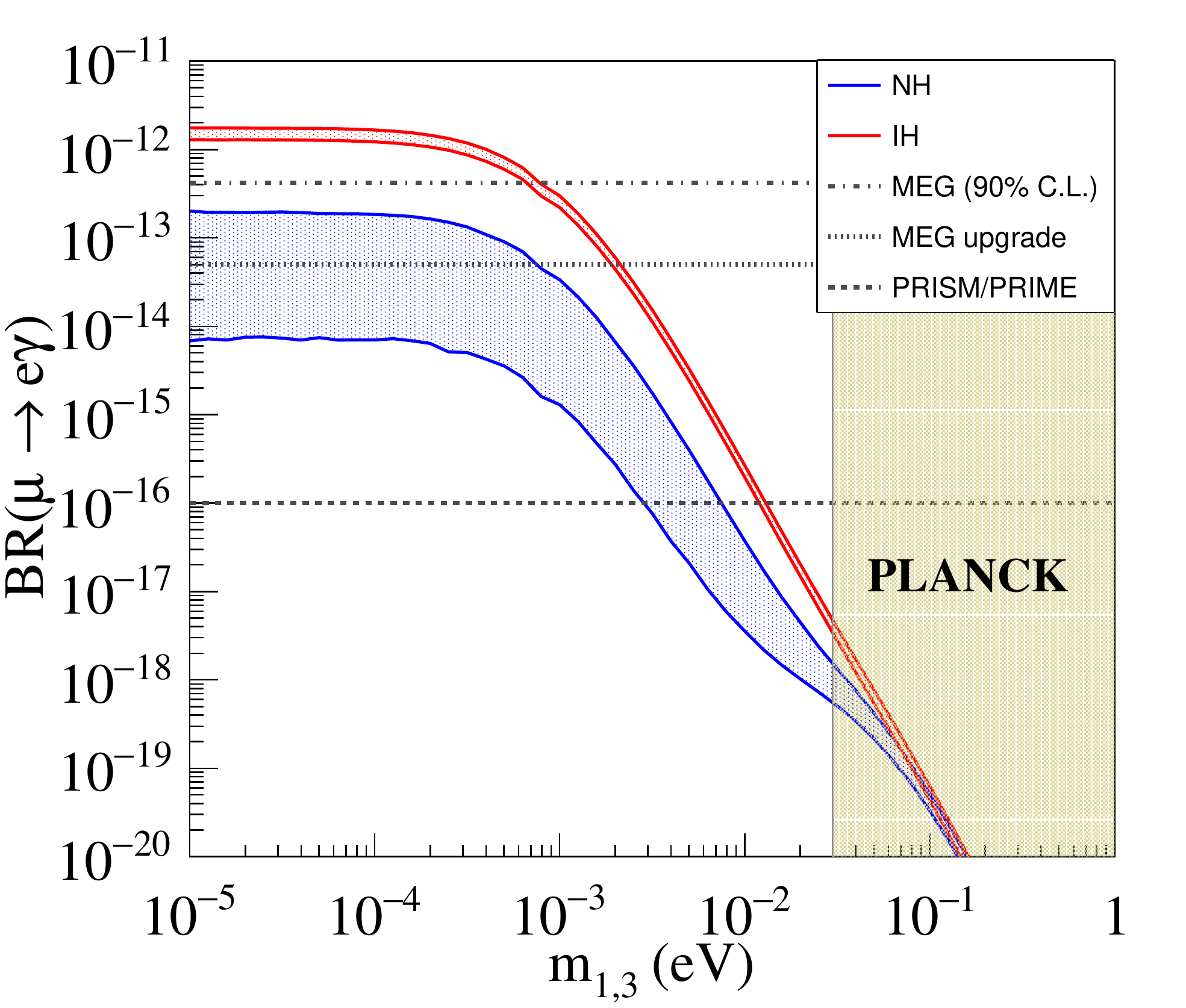}
	\includegraphics[width=0.38\textwidth]{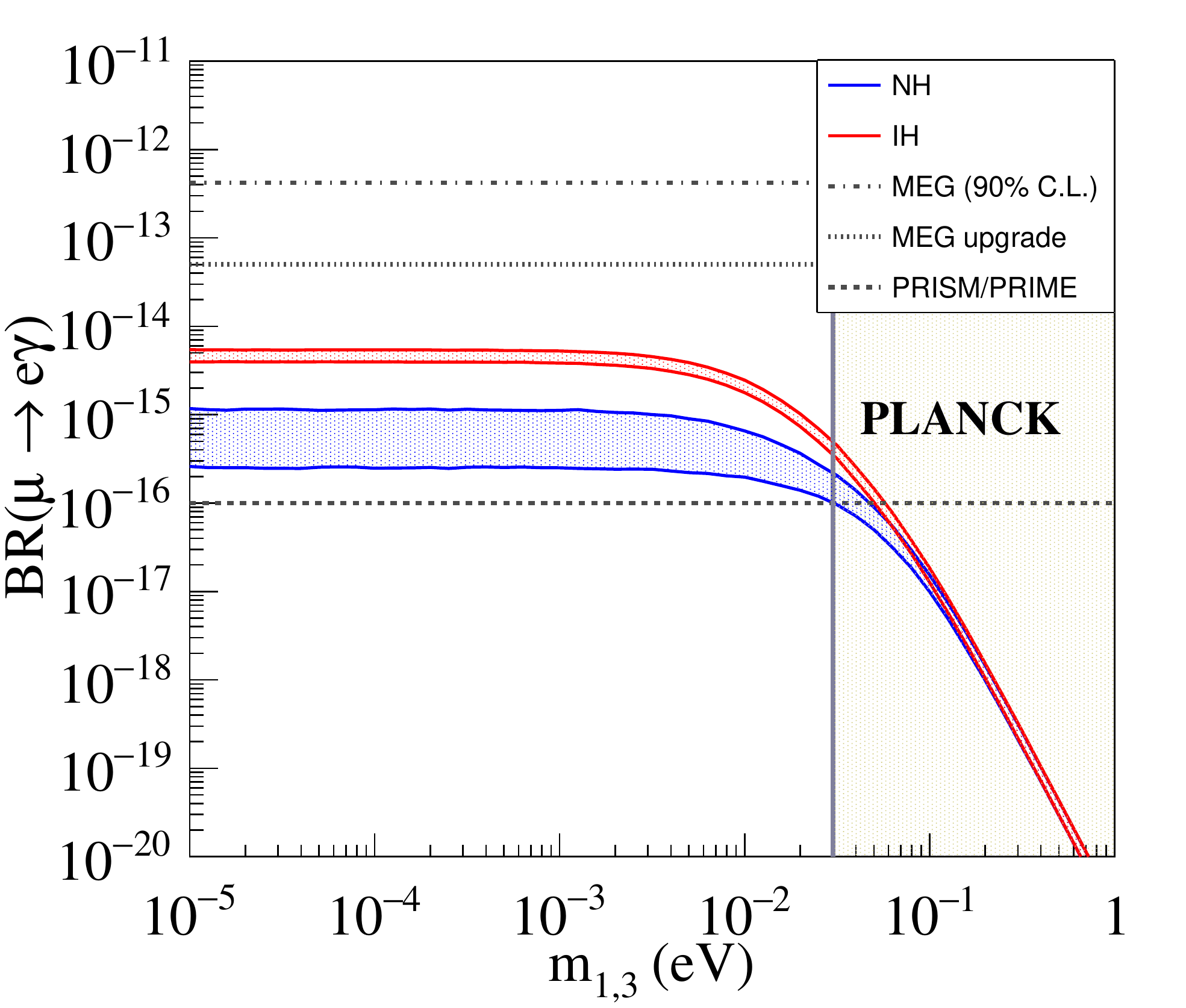}
	\caption{Branching ratio of the lepton flavor violating process, $\mu \to e \gamma$ as a function of lightest neutrino 
		mass $m_1$ for NH and $m_3$ for IH. The blue (NH) and red (IH) coloured regions display the model predictions on $\mu\rightarrow e \gamma$ due to the exchange of heavy right handed neutrino ($N_R$) (upper plot) and heavy sterile neutrino ($S_L$) (lower plot).}
	\label{fig:mutoegamma}
\end{figure}

The analytic expression for the branching ratio for the lepton flavor violating process $\muegam$ shown in Feynman diagram in Fig.\ref{fig:lfv-NS} due to mediation of heavy neutrinos ($N$ and $S$) is given by
\begin{equation}
\text{Br}_{\muegam} = \frac{\alpha_W^3s_W^2}{256\pi^2}\frac{m_\mu^4}{M_{W_L}^4}\frac{m_\mu}{\Gamma_\mu}|\mathcal{G}_\gamma^{\mu e}|^2\,, \label{mueg}
\end{equation}
where $\alpha_W=1/29.0$ is weak fine struture constant, $m_\mu=105$~MeV is the muon mass, $M_{W_L}$ is the SM $W$-boson mass, $s_W\equiv \sin\theta_W$ and $\Gamma_\mu=2.996\times 10^{-19}$ GeV is the total decay width of the muon. The important parameter,$G_\gamma^{\mu e}$, for deriving constraints on light neutrino masses, is of the following form, 
\begin{eqnarray}
G_\gamma^{\mu e}=\bigg| \sum_{i=1}^3 \bigg\{{\mbox{V}^{\nu N}_{\mu i}}^* {\mbox{V}^{\nu N}_{e i}}  \mathcal{G}_{\gamma} \left(x_{N_i}\right) 
+ {\mbox{V}^{\nu S}_{\mu i}}^* {\mbox{V}^{\nu S}_{e i}} \mathcal{G}_{\gamma} \left(x_{S_i}\right)  \bigg\}  \bigg|^2 
\label{eq:brmuegam_RR_full}
\label{gmue}
\end{eqnarray}
where $x_{N_i}=m_{N_i}^2/M_{W_L}^2$, $x_{S_i}=m_{S_i}^2/M_{W_L}^2$ and the form of loop function is given by
\begin{eqnarray}
G_\gamma(x) &=& -\frac{x(2x^2+5x-1)}{4(1-x)^3}-\frac{3x^3}{2(1-x)^4}\ln x \, 
\end{eqnarray}
The form of loop function $\mathcal{G}_\gamma (x)$ and its dependance with change of lightest neutrino mass $m_1$ (for NH) and $m_3$ (for IH) can be found in the appendix of the main manuscript. 
The other parameters $\mbox{V}^{\nu N}$ and $\mbox{V}^{\nu S}$ are mixing matrices representing mixing of light neutrinos with $N_R$ and $S_L$, respectively.

Fig.\ref{fig:mutoegamma} shows the variation of $\text{Br}_{\muegam}$ as a function of lightest neutrino mass  with contributions coming from purely $N_R$ (presented in the upper plot) and $S_L$ (presented in the lower plot). The blue color (NH) and red color (IH) regions are model prediction on $\mu \to e \gamma$ within $3-\sigma$ allowed range of neutrino mixing angles and mass squared differences. In x-axis, $m_1$ for NH ($m_3$ for IH) represents absolute value of light neutrino mass. The current experimental limit and future sensitivity by MEG ($\text{Br}_{\muegam} \leq 4.2\times 10^{-13}$) \cite{MEG:2013oxv}, MEG upgrade ($\text{Br}_{\muegam} \leq 5.0\times 10^{-14}$)~\cite{Baldini:2013ke} and PRISM/PRIME ($\text{Br}_{\muegam} \leq 1.0\times 10^{-16}$)~\cite{Kuno:2005mm} on branching ratio for $\mu \to e \gamma$ are presented in dashed horizontal lines. The vertical shaded regions are PLANCK bound on lightest neutrino mass with 95$\%$ C.L.~\cite{Tian:2020tur}. the experimental bounds are satisfied by the model predicted branching ratio and from these plots one can derive bound on lightest neutrino mass as follows. 

\begin{itemize}
	\item Saturating the MEG bound ($\text{Br}_{\muegam} \leq 4.2\times 10^{-13}$), one can derive the limit on lightest neutrino mass in the range of meV scale in IH case while most of the parameters are ruled out for NH case. 
	\item Saturating the MEG Upgrade bound  ($< 5.0 \times 10^{-14}\,$) one can find lightest neutrino mass less than few meV for NH case and $4$~meV for IH case. However, most of the paremeter spaces satisfying PRISM/PRIME ($<1.0 \times 10^{-16}\,$) bound are lying in the quasi-degenerate (QD) pattern of light neutrino masses which already ruled out by PLANCK data.
\end{itemize}

\section{LNV in LRSM}

We have also discussed how the light-heavy neutrino mixing leads to sizable new contributions to neutrinoless double beta decay in the model. We emphasize on left-handed current effects due to the exchange of heavy neutrinos $N_R$ and $S_L$.
\begin{figure}[htb!]
\centering
\includegraphics[scale=0.31]{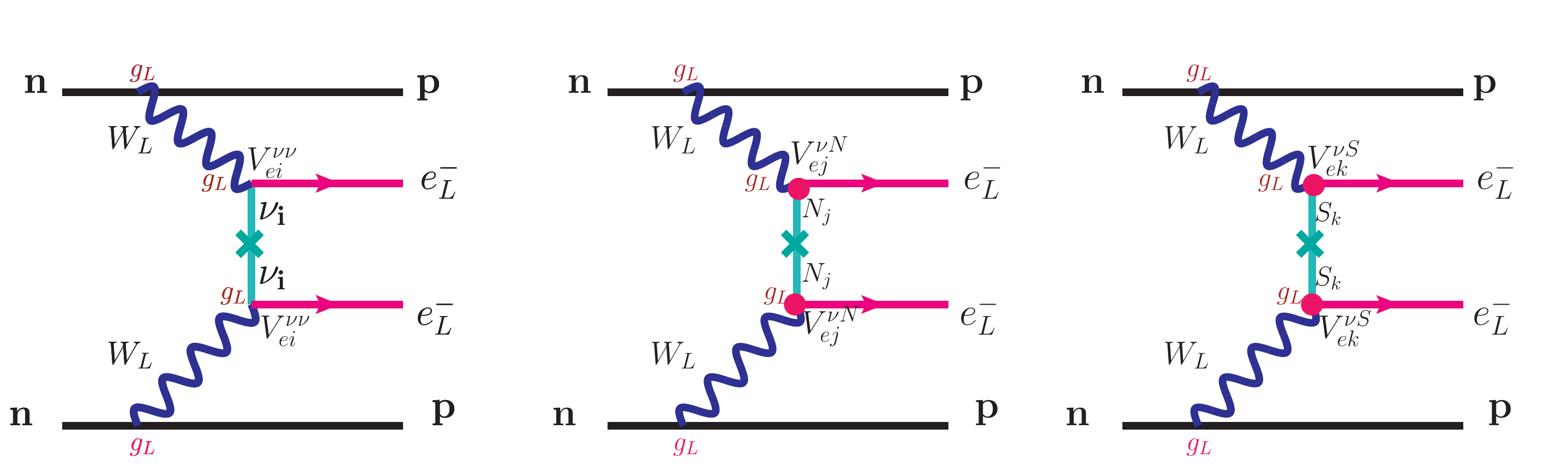}
 \caption{Feynman diagrams for $0\, \nu\, \beta \beta$ transition due to purely left-handed 
 currents with the exchange of light neutrino $\nu_i$ and heavy neutrinos $N_j$ and $S_k$ in flavour basis.}
\label{feyn:lrsm-WLL}
\end{figure}
 The inverse half-life of $0\nu\beta \beta$ transition for a given isotope in terms of Effective Majorana mass can be written as,
 \begin{eqnarray}
 \left[T_{1/2}^{0\nu}\right]^{-1} &=&  G^{0\nu}_{01} \bigg| \frac{{\cal M}^{0\nu}_\nu}{m_e} \bigg|^2   \bigg[\big|m^{\nu}_{ee} \big|^2 
               +  \big|m^{N}_{ee} \big|^2+ \big|m^{S}_{ee} \big|^2 \bigg] \nonumber \\
               &=&G^{0\nu}_{01} \bigg( \frac{{\cal M}^{0\nu}_\nu}{m_e}\bigg)^2 \cdot |m^{\rm eff}_{\beta \beta}|^2\,.
\label{eq:Hlife-b}
\end{eqnarray}
Here $G^{0\nu}$ is the phase-space factor, $\mathcal{M}^{0\nu}_\nu$ is the Nuclear Matrix Element and $|m^{\rm eff}_{\beta \beta}|^2$ is the sum of contributions from light active neutrinos $\nu_L$, heavy right-handed neutrinos $N_R$ and sterile neutrinos $S_L$.
\begin{table}[h]
 \centering
 \caption{$G^{0\nu}_{01}$ and NMEs~\cite{Meroni:2012qf}}
\vspace{5pt}
 \begin{tabular}{lcccc}
 \hline
 \\ {Isotope} & $G^{0\nu}_{01}$ $[{\rm yrs}^{-1}]$  
 & {${\cal M}^{0\nu}_\nu$} & {${\cal M}^{0\nu}_N$}  \\[2mm] 
\hline \\
$^{76}$Ge  & $5.77 \times 10^{-15}$  & 2.58--6.64 & 233--412  \\[1mm]
$^{136}$Xe  & $3.56 \times 10^{-14}$ & 1.57--3.85 & 164--172  \\[1mm] 
\hline
 \end{tabular}
  \label{tab:nucl-matrix}
\end{table}

\begin{figure}[h!]
\centering
\includegraphics[width=0.38\textwidth]{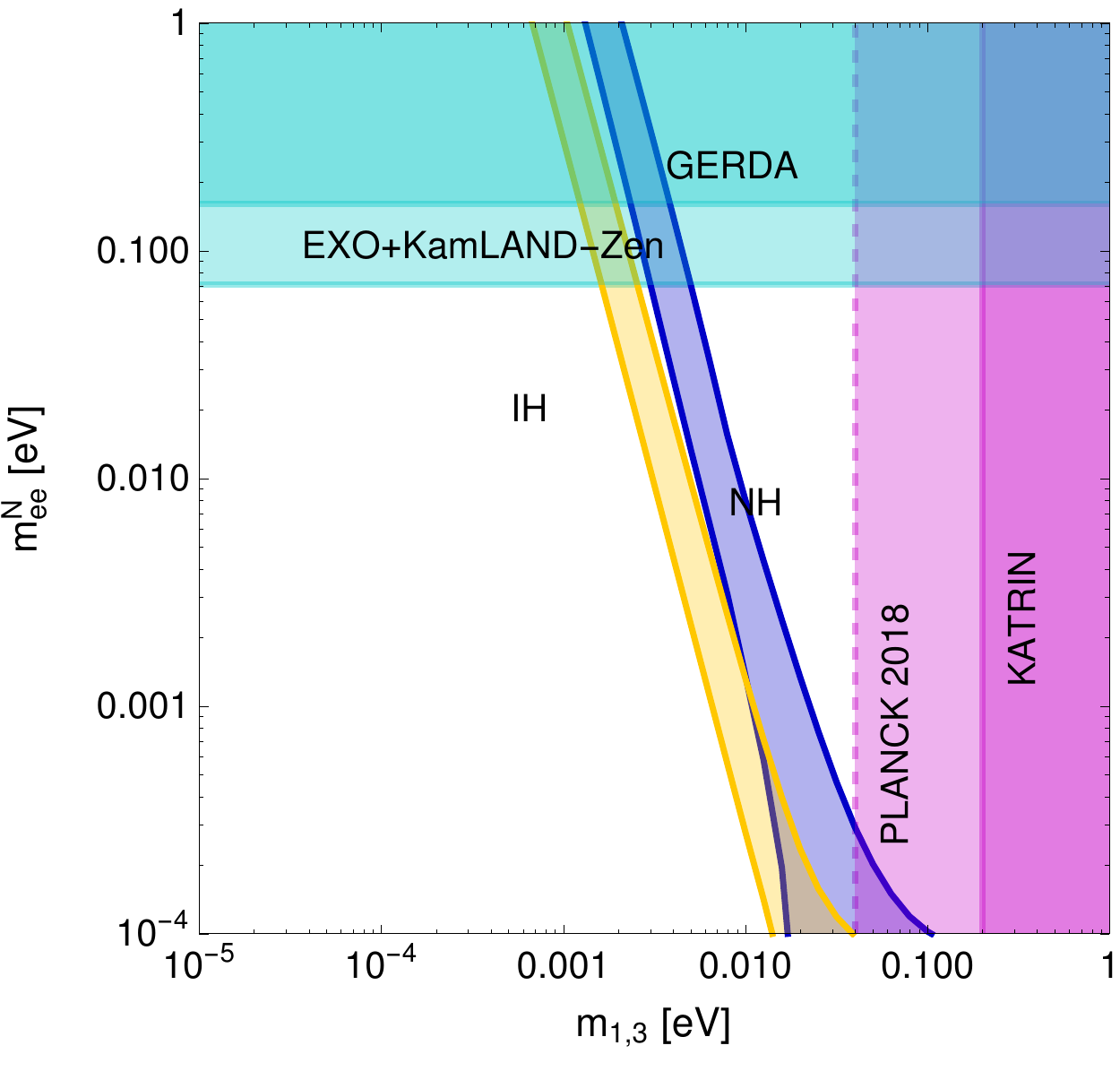}
\includegraphics[width=0.38\textwidth]{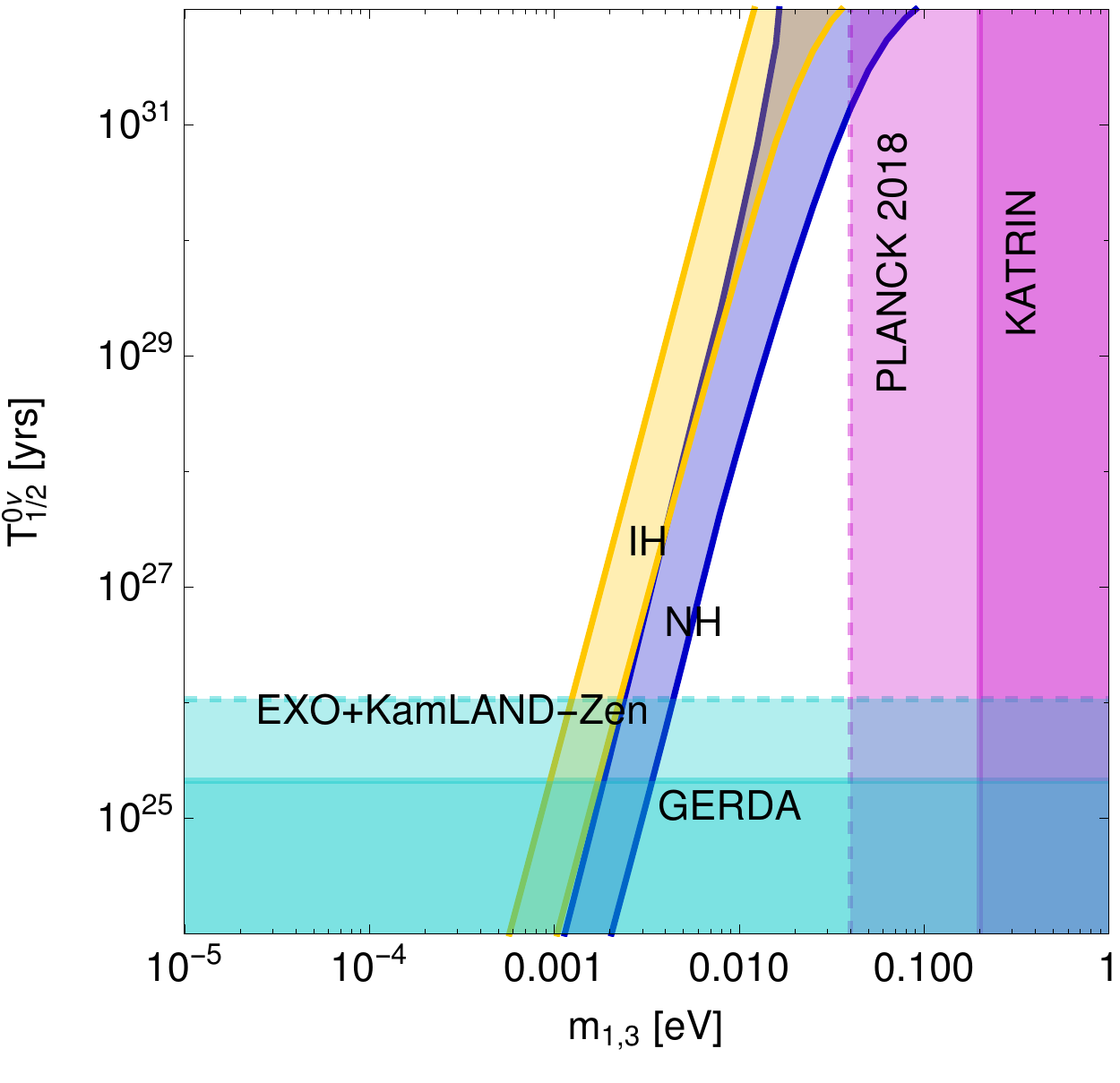}
\caption{Upper plot: New physics contribution to effective Majorana mass as a function of lightest neutrino mass, $m_1$ ($m_3$) for NH (IH) with the exchange of heavy neutrino $N_R$ and sterile neutrino $S_L$. Lower plot: Contributions of $N_R$ and $S_L$ to half-life vs lightest neutrino mass.}
\label{fig:0nubb-NS}
\end{figure}

\begin{table}[htp]
 \centering
 \caption{Limits on the half-life of $0\nu\beta\beta$.}
 \label{table:halflifelimits}
 \vspace{5pt}
 \begin{tabular}{lc}
  \hline
   Experiment & Limit \\[1mm]
  \hline HM & $1.9 \times 10^{25}\ {\rm yrs}$ \\
  GERDA & $2.1\times 10^{25}\ {\rm yrs}$ \\
  Combined $^{76}$Ge & $3.0\times 10^{25}\ {\rm yrs}$ \\
  GERDA Phase II & $5.2\times 10^{25}\ {\rm yrs}$ \\
  \hline EXO & $1.6\times 10^{25}\ {\rm yrs}$ \\
  KamLAND-Zen & $1.9\times 10^{25}\ {\rm yrs}$ \\
  Combined $^{136}$Xe & $3.4\times 10^{25}\ {\rm yrs}$ \\
  nEXO & $6.6\times 10^{27}\ {\rm yrs}$ \\
  \hline
 \end{tabular}
  \label{tab:half-life}
\end{table}

Fig.\ref{fig:0nubb-NS} shows the effective Majorana mass parameter in the upper plot and half-life in the lower plot due to the contributions of $N_R$ and $S_L$ as a function of lightest neutrino mass $m_1$(NH) and $m_3$(IH). The horizontal blue bands stand for the improved limits on $0\nu\beta\beta$ decay on effective Majorana mass and half-life with the combined results of GERDA~\cite{GERDA:2013vls} and KamLAND-Zen~\cite{KamLAND-Zen:2012mmx} experiments. The vertial magenta region is disfavoured by Planck-2018~\cite{Tian:2020tur} and KATRIN data~\cite{KATRIN:2019yun} on sum of light neutrino masses. The contributions of $N_R$ and $S_L$ to effective Majorana mass parameter and half-life predictions makes both NH(blue band) and IH(orange band) patterns of light neutrino masses sensitive to the current experimental bounds. In these plots the largest value of right-handed neutrino mass is fixed at $M_{N_R}=1$ TeV, the Majorana phases and Dirac CP-phase are varied between $0 \to \pi$, and other neutrino oscillation parameters are taken in their allowed $3\sigma$ range.
From these plots it is observed that, when the contributions of $N_R$ and $S_L$ are considered, the allowed values of lightest neutrino mass are found to be in the range of $10-25$~meV for $m_1$(NH) and $25-40$~meV for $m_3$(IH). This is done by saturating the effective Majorana mass with the current GERDA and KamLAND-Zen bounds. It remains the same for half-life.

\section{Correlation between $m_{ee}$ and sum of light neutrino masses $\sum m_i$}

\begin{figure}[h!]
\centering
\includegraphics[width=0.38\textwidth]{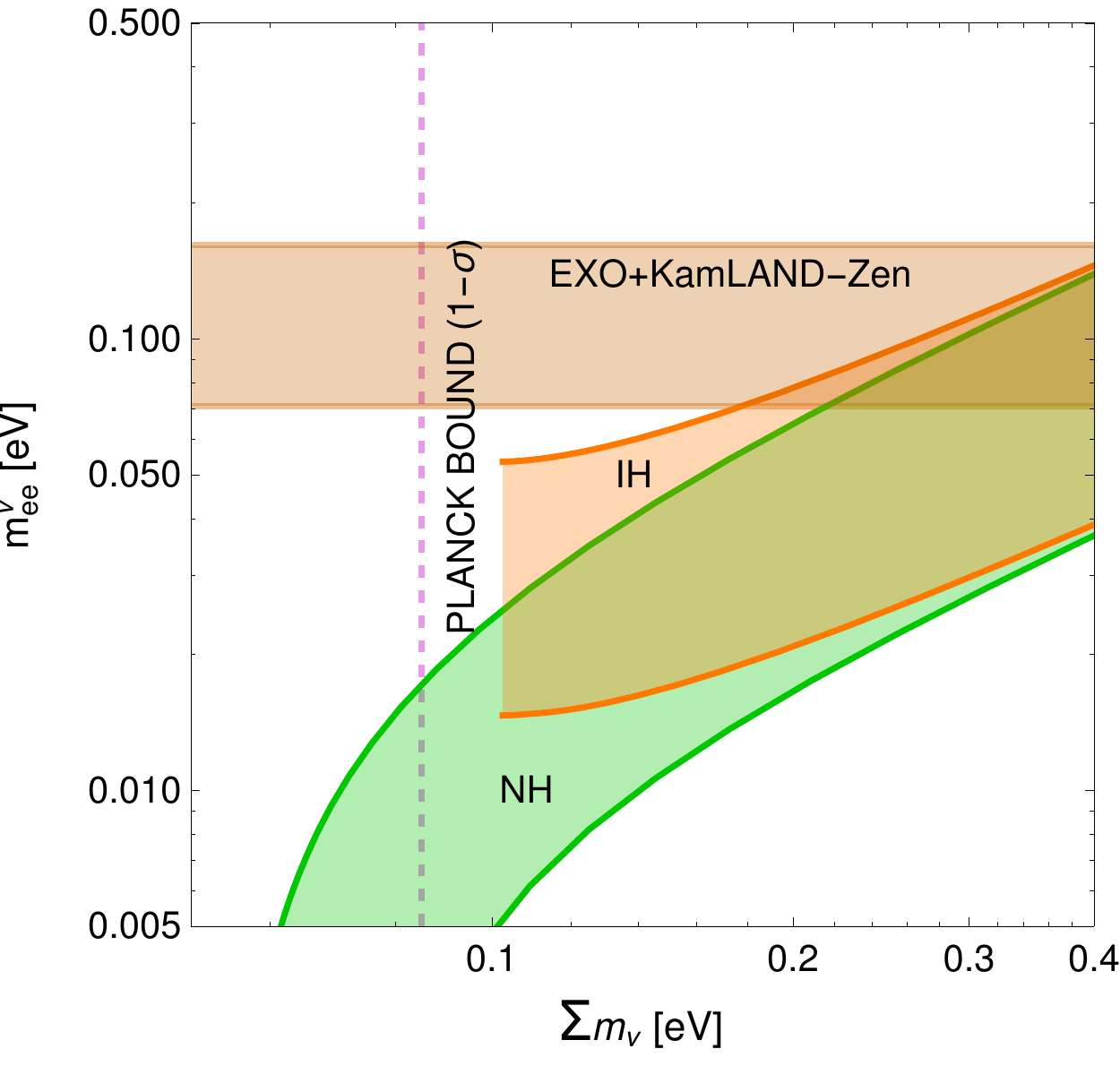}
\includegraphics[width=0.38\textwidth]{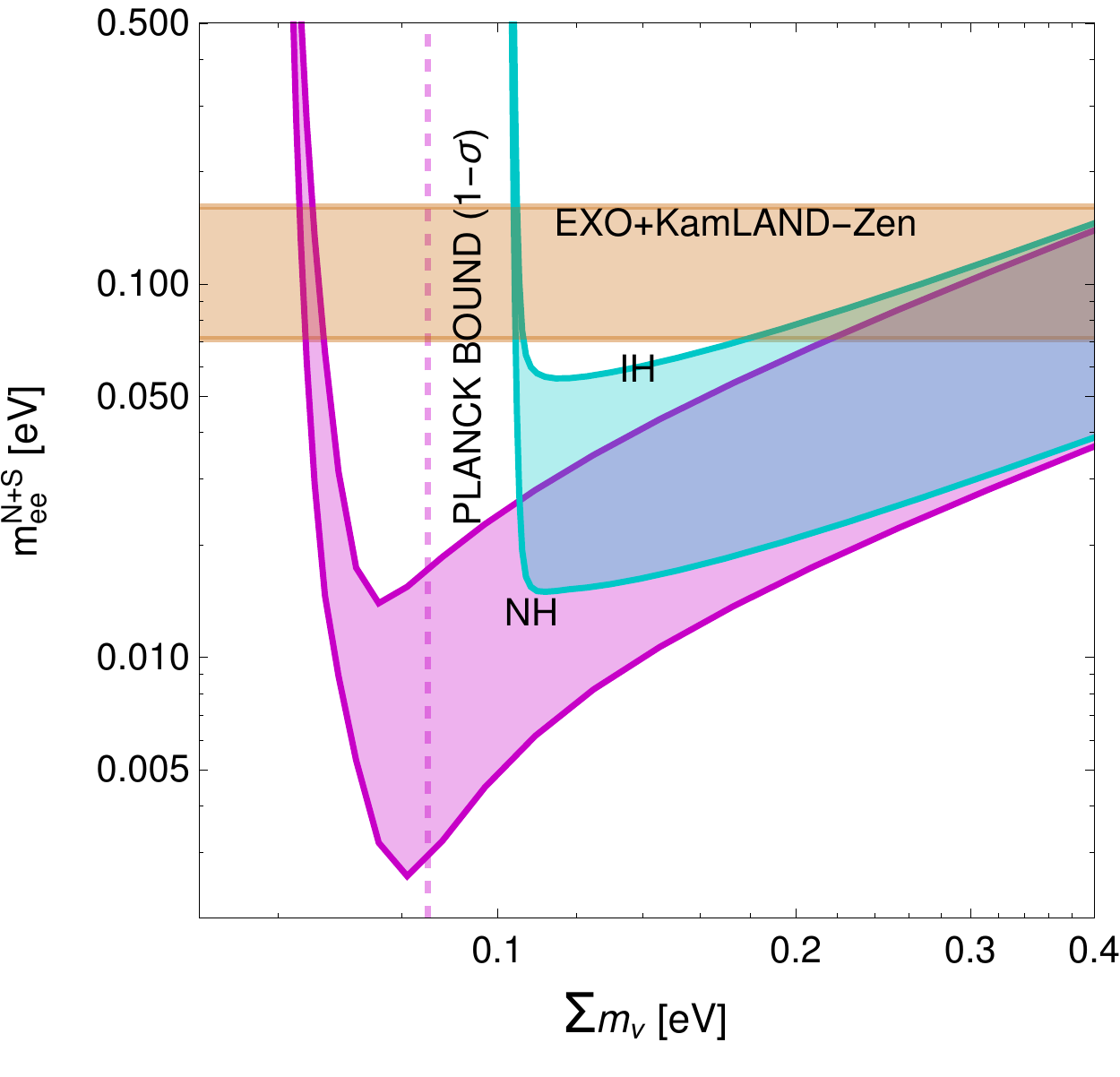}
\caption{Allowed region of effective Majorana mass parameter ($|m_{ee}|$) as a function of sum of light neutrino masses ($\Sigma m_i$) for standard mechanism (upper plot) and new contributions from $N_R$ and $S_L$ mediated diagrams (lower plot).}
\label{fig:mee-summass}
\end{figure}

We have also discussed the model predictions on effective Majorana mass by changing the sum of light neutrino masses $\Sigma m_i$. We use the following limits on sum of light neutrino masses~\cite{SDSS:2004kqt,Costanzi:2014tna,Palanque-Delabrouille:2014jca} for displaying 
the allowed region of $m_{ee}$ in Fig.\ref{fig:mee-summass}.
\begin{eqnarray}
&&m_\Sigma < \mbox{84\,meV}\quad \quad ~(1\sigma~\mbox{C.L.}) \nonumber \\
&&m_\Sigma < \mbox{146\,meV}\quad \quad (2\sigma~\mbox{C.L.}) \nonumber \\
&&m_\Sigma < \mbox{208\,meV}\quad \quad (3\sigma~\mbox{C.L.}) 
\end{eqnarray}

In both the plots of Fig.\ref{fig:mee-summass} the horizontal band represents the experimental bounds on effective Majorana mass $m_{ee}$ by EXO+KamLAND-Zen while the vertical dashed line shows the bound on sum of light neutrino masses from cosmology. The plot in the upper-panel shows the variation of effective Majorana mass parameter due to standard mechanism with sum of light neutrino masses. It shows that both NH (green band) and IH (red band) patterns of light neutrino masses are not sensitive to the current experimetal bounds on $m_{ee}$ and also disfavoured by cosmology. The plot in the lower panel shows the variation of effective Majorana mass parameter due to new physics contributions (contributions of $N_R$ and $S_L$) with sum of light neutrino masses. It shows that NH (purple band) pattern of light neutrino masses saturates the experimental bound on $m_{ee}$ as well as the cosmology bound on sum of light neutrino masses whereas IH (blue band) pattern is disfavoured by cosmology. Thus the model with type-II seesaw dominance gives an important result on the hierarchy of light neutrino masses when the contributions of heavy and sterile neutrinos are considered.

\section{Conclusion}
We considered the mechanism of natural type-II seesaw dominance that allows large light-heavy neutrino mixing in a TeV scale left-right symmetric model. This becomes possible with the addition of a sterile neutrino to the particle content. The model also generates large light-heavy neutrino mixing that gives new contributions to LFV process of $\mu \to e \gamma$ and neutrinoless double beta decay. As a result of these new contributions, the new branching ratios can be accessible to present as well as planned experiments. All the physical masses and mixing of neutral leptons are expressed in terms of light neutrino masses and PMNS mixing matrix. Thus the new physics contributions to LFV and LNV processes arising from heavy and sterile neutrinos depend upon light neutrino mass. As a result, the bound on absolute scale of light neutrino mass can be derived by saturating the experimental limits on these decay processes, which we find to be in the meV range. We also find that the new contributions to $0\nu\beta\beta$ decay can saturate the experimental bound for $m_{1} > 0.001$eV for NH pattern of light neutrino masses.



\bigskip 

\end{document}